\documentclass[journal,a4paper,twoside]{IEEEtran}

\usepackage{amssymb,epsfig,color}
\usepackage[cmex10]{amsmath}
\usepackage{graphicx}
\usepackage[font=footnotesize]{subfig}

\usepackage{flushend}

\makeatletter
\long\def\@makecaption#1#2{\ifx\@captype\@IEEEtablestring%
\footnotesize\begin{center}{\normalfont\footnotesize #1}\\
{\normalfont\footnotesize\scshape #2}\end{center}%
\@IEEEtablecaptionsepspace
\else
\@IEEEfigurecaptionsepspace
\setbox\@tempboxa\hbox{\normalfont\footnotesize {#1.}~~ #2}%
\ifdim \wd\@tempboxa >\hsize%
\setbox\@tempboxa\hbox{\normalfont\footnotesize {#1.}~~ }%
\parbox[t]{\hsize}{\normalfont\footnotesize \noindent\unhbox\@tempboxa#2}%
\else
\hbox to\hsize{\normalfont\footnotesize\hfil\box\@tempboxa\hfil}\fi\fi}
\makeatother

\makeatletter
\def\ps@IEEEtitlepagestyle{%
\def\@oddhead{\mbox{}\scriptsize\rightmark \hfil }%
\def\@evenhead{\scriptsize \hfil \leftmark\mbox{}}%
\def\@oddfoot{}%
\def\@evenfoot{}}
\makeatother

\newcommand{\abs}[1]{\vert #1 \vert}
\newcommand{\FF}{\vphantom{\vdots}}
\newcommand{\lan}{\langle}
\newcommand{\ran}{\rangle}

\begin{document}

\title{SNS: Analytic Receiver Analysis Software\\ Using Electrical Scattering Matrices}

%
\author{\IEEEauthorblockN{Oliver G. King\IEEEauthorrefmark{1}\IEEEauthorrefmark{2}}\\
\IEEEauthorblockA{\IEEEauthorrefmark{1}{\em California Institute of Technology, MC 249-17, 1200 E. California Blvd., Pasadena, CA, 91125}}\\
\IEEEauthorblockA{\IEEEauthorrefmark{2}Contact: ogk@astro.caltech.edu}
}


\maketitle

\begin{abstract}
SNS is a MATLAB-based software library written to aid in the design and analysis of receiver architectures. It uses electrical scattering matrices and noise wave vectors to describe receiver architectures of arbitrary topology and complexity. It differs from existing freely-available software mainly in that the scattering matrices used to describe the receiver and its components are analytic rather than numeric. This allows different types of modeling and analysis of receivers to be performed.

Non-ideal behavior of receiver components can be parameterized in their scattering matrices. \textbf{SNS} enables the instrument designer to then derive analytic expressions for the signal and noise at the receiver outputs in terms of parameterized component imperfections, and predict their contribution to receiver systematic errors precisely. This can drive the receiver design process by, for instance, allowing the instrument designer to identify which component imperfections contribute most to receiver systematic errors, and hence place firm specifications on individual components. Using \textbf{SNS} to perform this analysis is preferable to traditional Jones matrix-based analysis as it includes internal reflections and is able to model noise: two effects which Jones matrix analysis is unable to describe.

\textbf{SNS} can be used to model any receiver in which the components can be described by scattering matrices. Of particular interest to the sub-mm and terahertz frequency regime is the choice between coherent and direct detection technologies. Steady improvements in mm and sub-mm Low Noise Amplifiers (LNAs) mean that coherent receivers with LNAs as their first active element are becoming increasingly competitive, in terms of sensitivity, with bolometer-based receivers at frequencies above $\sim 100$~GHz.

As an example of the utility of \textbf{SNS}, we use it to compare two polarimeter architectures commonly used to perform measurements of the polarized Cosmic Microwave Background: differencing polarimeters, an architecture commonly used in polarization sensitive bolometer-based polarimeters; and pseudo-correlation polarimeters, an architecture commonly used in coherent, LNA-based, polarimeters. We parameterize common sources of receiver systematic errors in both architectures and compare them through their Mueller matrices, which encode how well the instruments measure the Stokes parameters of the incident radiation. These analytic Mueller matrices are used to demonstrate the different sources of systematic errors in differencing and correlation polarimeters.
\end{abstract}

\section{Introduction}

Many fields of astrophysics aim to measure increasingly faint signals. For instance, there is great interest at present in detecting and characterizing the B-mode \cite{Zaldarriaga:1997p247} of the polarized Cosmic Microwave Background (CMB). The strength of this signal is not yet determined by theory, but a strong upper limit is 170~nK \cite{Page:2007p205}, a tiny fraction of the CMB total intensity signal ($\sim2.7$~K).

An instrument built to detect very faint signals will almost certainly be heavily affected by systematic errors. It is increasingly important to be able to model the effects of receiver systematic errors on the measured signal, and on the receiver sensitivity. We want to be able predict the level of receiver systematic errors, show their impact on the gathered data, and make quantitative comparisons between different receiver architectures.

Previous analytic and semi-analytic approaches to characterizing systematic effects in receivers have usually employed Jones matrices to describe receiver components and Mueller matrices to characterize the effects of receiver systematics on the observed Stokes parameters \cite{Hu:2003p1726, ODea:2007p1678}. The use of Jones matrices to describe individual receiver components has several shortcomings. Only the forward path of the signal through the instrument is modeled -- internal scattering caused by reflections from poorly matched components is not included; and Jones matrix modeling is unable to describe component noise, and hence receiver sensitivity. Modeling a receiver with a full analytic description of the outputs and sensitivity in terms of individual component parameters allows us to identify which parameters of each component in a receiver are most important, and concentrate our efforts on improving them.

This paper introduces a technique and software for developing full analytic descriptions of receiver outputs and sensitivities in terms of lab-measureable errors in individual components. In this technique components are modeled by electrical scattering matrices. When describing a network of components with Jones matrices the forward-path cascaded response can be obtained through simple matrix manipulation and multiplication. The scattering matrix formulation does not share this simplicity of calculation: only the case of cascaded 2-port devices is amenable to a relatively simple analytic solution. This paper describes an algorithm for calculating the response of arbitrarily connected networks of components. We present software which implements this algorithm, and apply it to two common polarimeter architectures: differencing polarimeters, and pseudo-correlation polarimeters.

This software allows us to make robust analytic comparisons of receiver architectures. Errors in individual receiver components can be parameterized and propagated into the description of the receiver performance, e.g. the instrument Mueller matrix. We hence have a powerful tool for guiding the instrument design process and diagnosing the causes of non-ideal instrument behavior.

\section{Electrical Scattering Matrices}

We model the behavior of individual receiver components, and the full receiver, using electrical scattering matrices. The electrical scattering matrix (hereafter referred to as the scattering matrix) is a representation of a network using the ideas of incident, reflected, and transmitted waves. It provides a complete description of an $N$-port network as seen at its $N$ ports. A significant advantage of modeling receiver components with scattering matrices is that \emph{noise} can easily be included in the modeling. The noise produced by a device is modeled with a noise wave vector; see e.g. \cite{Wedge:1992p149}.

Consider the arbitrary $N$-port network shown in Figure~\ref{fig:N-port_network}. We denote the incident wave at port $i$ by $V_{i}^{+}$, the reflected wave by $V_{i}^{-}$, and the noise wave produced by the network at that port by $c_{i}$. These quantities are related by the scattering matrix $\mathbf{S}$ and noise wave vector $\mathbf{c}$ as follows:
\begin{equation}
\renewcommand{\arraystretch}{1.4}
\begin{bmatrix} 
V_{1}^{-}  \\  
\FF V_{2}^{-}  \\ 
\FF \vdots  \\ 
\FF V_{N}^{-}
\end{bmatrix}
= 
\begin{bmatrix} 
S_{11} & S_{12} & \cdots & S_{1N}  \\  
\FF S_{21} & 		       &  	      & \vdots         \\ 
\FF \vdots        &   		       &  	      &            \\ 
\FF S_{N1} & \cdots        &  	      & S_{NN} 
\end{bmatrix}
\begin{bmatrix} 
V_{1}^{+}  \\  
\FF V_{2}^{+}  \\ 
\FF \vdots  \\ 
\FF V_{N}^{+}
\end{bmatrix}
+
\begin{bmatrix} 
c_{1}  \\  
\FF c_{2}  \\ 
\FF \vdots  \\ 
\FF c_{N}
\end{bmatrix}
\label{eqn:scattering_matrix_description}
\end{equation}

The noise wave voltages $c_i$ of an $N$-port network are complex time-varying random variables characterized by a correlation matrix $\mathbf{C}$
\begin{align*}
\mathbf{C} = & \lan \mathbf{c} \otimes \mathbf{c}^{\dagger} \ran \\
= &
\begin{bmatrix} 
\lan \vert c_1 \vert^{2} \ran  & \lan c_1 c_2^* \ran  & \cdots & \lan c_1 c_N^* \ran  \\  
\FF \lan c_2 c_1^* \ran & 		       &  	      & \vdots         \\ 
\FF \vdots        &   		       &  	      &            \\ 
\FF \lan c_N c_1^* \ran & \cdots        &  	      & \lan \vert c_N \vert^{2} \ran
\end{bmatrix}
\end{align*}
where the angle brackets indicate time averaging, $\dagger$ indicates the conjugate transpose operation, and $\otimes$ indicates the outer product (or Kronecker product). The diagonal terms of $\mathbf{C}$ give the noise power deliverable at each port in a 1~Hz bandwidth. The off-diagonal terms are correlation products. The noise correlation matrix $\mathbf{C}$ for a passive network is determined from its scattering matrix $\mathbf{S}$ by \cite{Wedge:1991p159}
\begin{equation}
 \mathbf{C} = kT(\mathbf{I}-\mathbf{S}\mathbf{S}^{\dagger}) \label{eqn:noise_correlation_matrix_for_passive_device}
\end{equation}
where $k$ is Boltzmann's constant, $T$ is the physical temperature of the network, and $\mathbf{I}$ is the identity matrix. The noise correlation matrix for an active network can be determined by measurement or modeling.

\section{Solving Arbitrary Networks}

\begin{figure}
\centering
\subfloat[]{
 \includegraphics[width=3.35in]{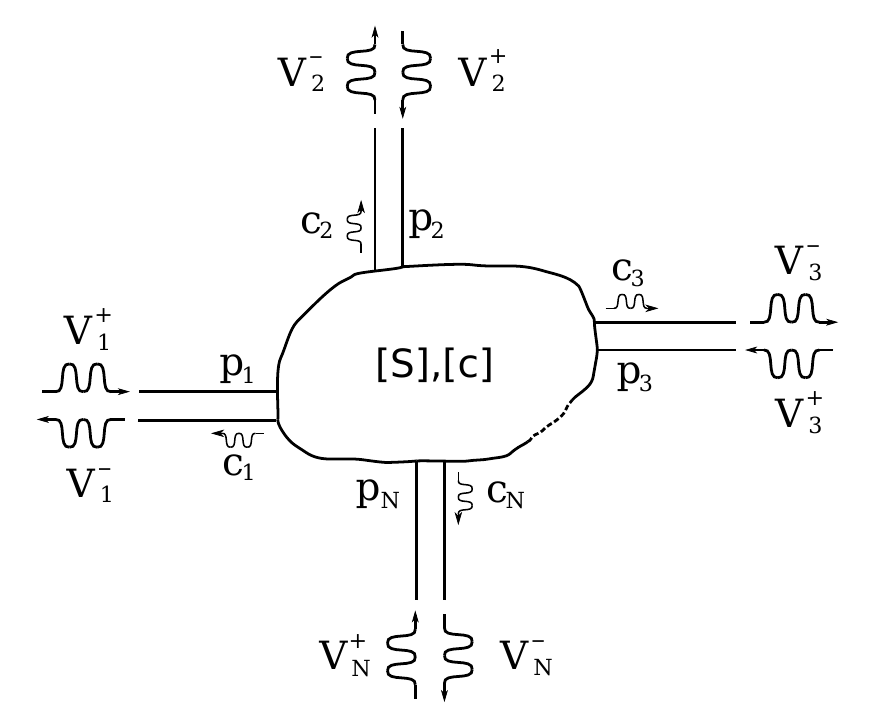}
 \label{fig:N-port_network}
}
\\
\subfloat[]{
 \includegraphics[width=2in]{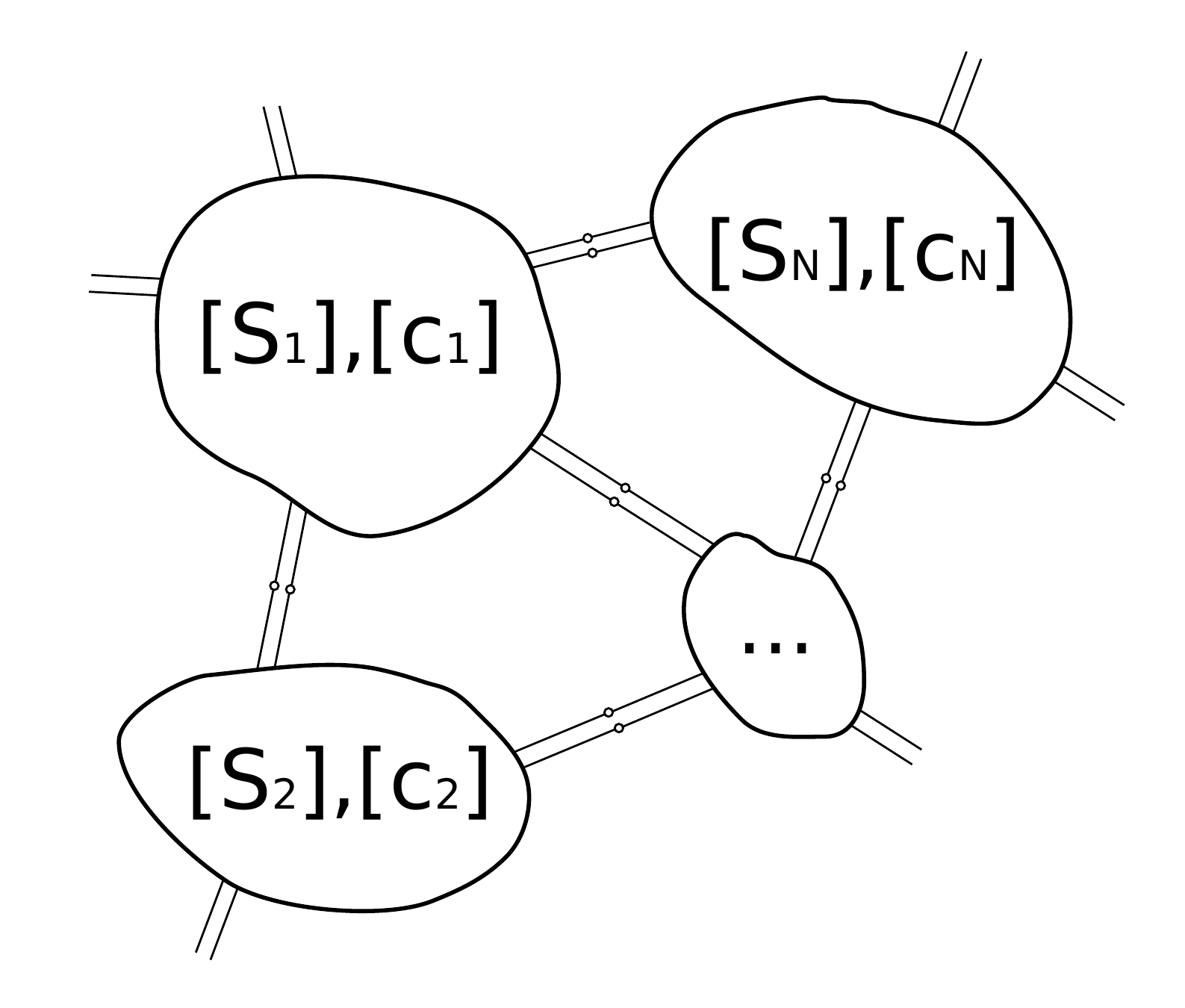}
 \label{fig:arbitrary_N_port_network}
}
\caption{(a) An arbitrary $N$-port network. The total signal at each port $p_i$ consists of an incident signal $V_i^+$, a reflected signal $V_i^-$, and a noise signal $c_i$. The reflected signal $V_i^-$ is a weighted sum of the incident signal at port $i$ and transmitted signals from the other ports of the network, the coefficients of the sum being the elements of the network scattering matrix $\mathbf{S}$. The noise signals are given by the noise wave vector $\mathbf{c}$. (b) An arbitrarily connected network of N-port devices.}
\label{anotherlabel}
\end{figure}

Consider the arbitrarily connected network of $N$-port networks shown in Figure~\ref{fig:arbitrary_N_port_network}. We need an algorithm to calculate the scattering matrix $\mathbf{S}$ and noise wave vector $\mathbf{c}$ which describe the connected network. The algorithm derived here is an extension of the algorithm described in \cite{Filipsson:1981p60}, with added noise wave vector manipulation. Similar algorithms are used numerically in {\verb SUPERMIX } \cite{Ward:1999p26}.

First, let us consider the effect of connecting together ports $k$ and $m$ of an $N$-port network described by Equation~\ref{eqn:scattering_matrix_description}. Connecting the ports means that $V_{k}^{+} = V_{m}^{-}$ and $V_{k}^{-} = V_{m}^{+}$. Manipulation of rows $k$ and $m$ of Equation~\ref{eqn:scattering_matrix_description} then gives us the expressions
\begin{align}
V_m^- & = \sum_{i\neq k,m} \frac{S_{mi}}{1-S_{mk}}V_{i}^{+} + \frac{S_{mm}}{1-S_{mk}}V_{k}^{-} + \frac{c_{m}}{1-S_{mk}} \label{eqn:Vmminus_expression}\\
V_k^- & = \sum_{i\neq k,m} \frac{S_{ki}}{1-S_{km}}V_{i}^{+} + \frac{S_{kk}}{1-S_{km}}V_{m}^{-} + \frac{c_{k}}{1-S_{km}} \label{eqn:Vkminus_expression}
\end{align}

By substituting Equations~\ref{eqn:Vmminus_expression} and \ref{eqn:Vkminus_expression} into each other we can obtain expressions for $V_{m}^{-}$ and $V_k^{-}$. Substituting these expressions into Equation~\ref{eqn:scattering_matrix_description} we can obtain a new expression for the reflected wave $V_{i}^{-}$:
\begin{align}
\nonumber V_i^- = & \sum_{j\neq k,m} \Bigg[ S_{ij} + \frac{(1-S_{km})(1-S_{mk})}{(1-S_{km})(1-S_{mk})-S_{kk}S_{mm}} \\
\nonumber & \Bigg( \frac{S_{ik}S_{mj}}{1-S_{mk}} + \frac{S_{ik}S_{mm}S_{kj}}{(1-S_{mk})(1-S_{km})}  \\
\nonumber & + \frac{S_{im}S_{kj}}{1-S_{km}} + \frac{S_{im}S_{kk}S_{mj}}{(1-S_{km})(1-S_{mk})}\Bigg) \Bigg]V_j^+  \\
\nonumber & + c_i + S_{ik} \Bigg( \frac{(1-S_{km})c_{m}+S_{mm}c_{k}}{(1-S_{km})(1-S_{mk})-S_{kk}S_{mm}}\Bigg) \\
		 &  + S_{im} \Bigg( \frac{(1-S_{mk})c_{k}+S_{kk}c_{m}}{(1-S_{km})(1-S_{mk})-S_{kk}S_{mm}}\Bigg) \label{eqn:new_Viminus_expression}
\end{align}

From Equation~\ref{eqn:new_Viminus_expression} we obtain replacement expressions for the elements $S_{ij}$ of $\mathbf{S}$ and the noise waves $c_i$:
\begin{align}
\nonumber S_{ij}^{\textrm{new}} = & S_{ij} + A \Big[S_{ik}S_{mj}(1-S_{km}) + S_{ik}S_{kj}S_{mm} \\
  & + S_{im}S_{kj}(1-S_{mk})+S_{im}S_{mj}S_{kk}\Big] \label{eqn:S_replacement_formula} \\
\nonumber c_{i}^{\textrm{new}} = & c_{i} + A  \Big[ \big(S_{im}S_{kk}+S_{ik}(1-S_{km})\big)c_{m} \\ 
 & +\big(S_{ik}S_{mm}+S_{im}(1-S_{mk})\big)c_{k}\Big] \label{eqn:c_replacement_formula} \\
\nonumber \textrm{where } A = & \frac{1}{(1-S_{km})(1-S_{mk})-S_{kk}S_{mm}}
\end{align}

Rows and columns $k$ and $m$ are then removed from $\mathbf{S}$ and rows $k$ and $m$ are removed from $\mathbf{c}$ to create the scattering matrix and noise vector which describe the new $(N-2)$-port network.

\subsection{Algorithm}\label{par:network_evaluation_algorithm}

To obtain the scattering matrix and noise wave vector which describe the arbitrarily connected network shown in Figure~\ref{fig:arbitrary_N_port_network} begin by forming the scattering matrix and noise wave vector which describe the unconnected network:
\begin{align}
\mathbf{S}
=
\begin{bmatrix} 
\mathbf{S}_1  & \mathbf{0}  & \cdots & \mathbf{0}  \\  
\FF \mathbf{0} & \mathbf{S}_2	       &  	      & \vdots         \\ 
\FF \vdots        &   		       &  	      &            \\ 
\FF \mathbf{0} & \cdots        &  	      & \mathbf{S}_N
\end{bmatrix}
,
\mathbf{c}
=
\begin{bmatrix} 
\mathbf{c}_1  \\  
\FF \mathbf{c}_2         \\ 
\FF \vdots    \\ 
\FF \mathbf{c}_N
\end{bmatrix}
\label{eqn:scattering_of_unconnected_network}
\end{align}

We then successively form each connection in the network. For each connection, find the rows and columns $k$ and $m$ of $\mathbf{S}$ and $\mathbf{c}$ in Equation~\ref{eqn:scattering_of_unconnected_network} which correspond to the ports being connected. Use the replacement formulae given by Equations~\ref{eqn:S_replacement_formula} and \ref{eqn:c_replacement_formula} to adjust the $\mathbf{S}$ matrix and $\mathbf{c}$ vector. Remove rows and columns $k$ and $m$ from $\mathbf{S}$, and rows $k$ and $m$ from $\mathbf{c}$. Repeat for each remaining connection until we are left with the scattering matrix $\mathbf{S}$ and noise vector $\mathbf{c}$ which describe the fully connected network.

\section{Software Implementation} \label{sec:software_implementation}

We have derived an algorithm in \S\ref{par:network_evaluation_algorithm} for finding the scattering matrix and noise wave vector which describe an arbitrarily connected network. We need to be able to apply it to arbitrary receivers with parameterized scattering matrices describing the receiver components and obtain analytic expressions for the outputs and noise in terms of the component parameters.

The algorithm must be implemented in a programming language with the ability to manipulate symbolic algebraic expressions. This programming language must also support pointers (or equivalent data structure) to allow the creation of a navigateable network. We implemented the algorithm in MATLAB\footnote{http://www.mathworks.com/}, which has a powerful and well developed symbolic algebra toolbox. While it does not have a native pointer data type (as of version R2008b), a third party open-source pointer library\footnote{http://code.google.com/p/pointer/} adds this capability. The software package we developed to perform this modeling is called {\verb SNS }\footnote{Download at http://www.astro.caltech.edu/$\sim$ogk/SNS/}.

\subsection{Representing a Network}

\begin{figure}
 \centering
 \includegraphics[width=3.25in]{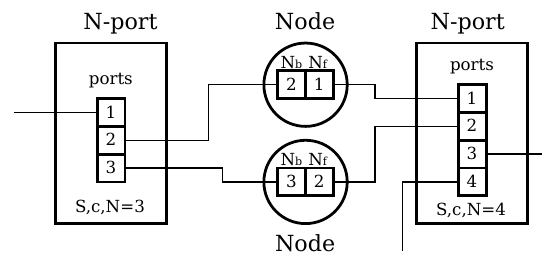}
 \caption{Schematic showing the nature of node and $N$-port objects and how they connect to each other in the software.}
 \label{fig:representing_a_network}
\end{figure}

A network is represented by nodes and $N$-port objects, as shown in Figure~\ref{fig:representing_a_network}. They are both pointer objects. Each $N$-port object contains an array of references to the nodes connected to each of its $N$ ports, a scattering matrix $\mathbf{S}$, a noise wave vector $\mathbf{c}$, and a variable $N$, the number of ports of the object.

Each node object contains two pointers; these refer to the $N$-port objects the node connects to in the ``forward'' and ``backward'' directions, and which port number the connection is made to ($N_{fp}$ and $N_{bp}$ respectively). Note that the forward and backward directions are completely arbitrary; they are merely a helpful concept when trying to visualize the operation of the algorithms which act on the network.

The network is constructed by creating all the node and $N$-port objects using functions called {\verb makeNode() } and {\verb makeNport() }, assigning scattering matrices and noise wave vectors to the $N$-port objects, and connecting each node to its forward and backward $N$-port objects. A {\verb connectNode } function hides the complexity of assigning references to the appropriate array locations and assigning the appropriate port numbers to variables.

Nodes are classified into four types: input, output, central, and terminated. We want to calculate the performance of a network in terms of the response seen at the outputs due to signals presented at the inputs. The central and terminated nodes are removed by the network-solving program.

Once all the objects have been created, assigned matrices and vectors, and connected, it suffices to describe the network by the four arrays of nodes. Due to the fully connected nature of the network representation it is possible to start at any node and navigate to any other node by following the appropriate links between objects.

\subsection{Solving a Network} \label{sec:solving_a_network}

The network-solving program accepts four arrays of nodes: the inputs, outputs, central nodes, and terminated nodes. It returns the scattering matrix and noise wave vector for the connected network, where the central and terminated nodes have been removed through application of the algorithm described in \S\ref{par:network_evaluation_algorithm}.

The first step the software performs is to remove the terminated nodes, if there are any. The software assumes that the terminations are perfectly matched and at a common physical temperature. It modifies the object scattering matrices to remove the terminated nodes, and adds the noise terms produced by the terminated nodes to the noise wave vectors.

The network \emph{sans} terminated nodes is then passed to a recursive network shrinking program. This program begins with the first central node and applies the algorithm given by Equations~\ref{eqn:S_replacement_formula} and \ref{eqn:c_replacement_formula} to a sub-network consisting of the two $N$-port objects connected to that particular node. A new $N$-port object is created and assigned the resulting scattering matrix and noise wave vector. All the nodes which were connected to the now-defunct $N$-port objects are reconnected to this new $N$-port object at the appropriate ports. A new network is formed by excluding the central node just considered and the program is recursively called on this new network. This continues until there are no more central nodes, at which point the scattering matrix and noise wave vector of the single remaining $N$-port object are returned.

Applying the algorithm in this fashion, rather than to the entire network at once, means that the size of the matrix $\mathbf{S}$ in Equation~\ref{eqn:scattering_of_unconnected_network} is kept small, speeding up computation. This is not an optimum solution to network shrinking, but we have found it to be significantly faster than applying the algorithm to the full unconnected network for networks of more than a few $N$-port objects.

The description given above glosses over the significant complexity in keeping track of which nodes should be connected where, and certain configurations of nodes and $N$-port objects which would cause the default implementation of the algorithm to fail. The majority of the code is dedicated to performing these functions; only a small fraction of the code actually carries out the calculations described by the algorithm.

\subsection{Example Code Listing}

\begin{figure}[!t]
 \includegraphics[width=3.49in]{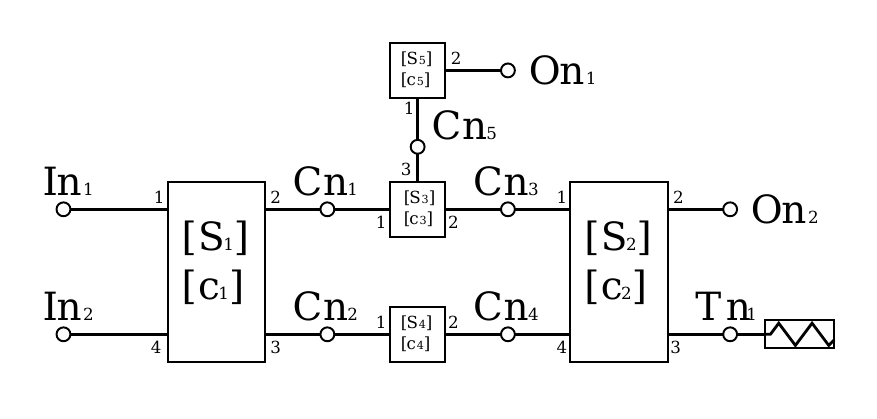}
\caption{An arbitrary network of $N$-port devices to illustrate the software. Nodes are indicated by open circles, $N$-port devices by rectangles. $N$-port device $i$ is described by scattering matrix $\mathbf{S}_{i}$ and noise wave vector $\mathbf{c}_{i}$. Input nodes to the network are $\textrm{In}_{i}$, central nodes $\textrm{Cn}_{i}$, terminated nodes $\textrm{Tn}_{i}$, and output nodes $\textrm{On}_{i}$.}
 \label{fig:software_example_network}
\end{figure}

To illustrate the operation of the program consider the network shown in Figure~\ref{fig:software_example_network}. It is represented in software by connected lists of nodes and $N$-port objects, as shown in the following code listing:
{\tiny
\begin{verbatim}
% Matrices S1, S2, S3, S4, S5 and vectors c1, c2, c3, c4, c5 assumed to 
% have been previously defined using symbolic algebra library.
% Make the N-port objects
P1 = makeNport(); P2 = makeNport(); P3 = makeNport();
P4 = makeNport(); P5 = makeNport();
% Assign scattering matrices and noise vectors
P1.S = S1; P2.S = S2; P3.S = S3; P4.S = S4; P5.S = S5;
P1.c = c1; P2.c = c2; P3.c = c3; P4.c = c4; P5.c = c5;
% Make the nodes which connect the N-port objects
In1 = makeNode(); In2 = makeNode(); On1 = makeNode();
On2 = makeNode(); Tn1 = makeNode(); Cn1 = makeNode(); 
Cn2 = makeNode(); Cn3 = makeNode();
Cn4 = makeNode(); Cn5 = makeNode();
% Connect nodes to N-ports
connectNode(In1,[],1,P1,1); connectNode(In2,[],1,P1,4);
connectNode(Cn1,P1,2,P3,1); connectNode(Cn2,P1,3,P4,1);
connectNode(Cn3,P3,2,P2,1); connectNode(Cn4,P4,2,P2,4); 
connectNode(Cn5,P3,3,P5,1); connectNode(Tn1,P2,3,[],1); 
connectNode(On1,P5,2,[],1); connectNode(On2,P2,2,[],1);
% Build arrays of nodes
inputs = {In1 In2}; outputs = {On1 On2}; 
cnodes = {Cn1 Cn2 Cn3 Cn4 Cn5}; tnodes = {Tn1};
% Pass arrays of nodes to network calculator
[S, c] = getScatteringRecursive(inputs,outputs,cnodes,tnodes);
\end{verbatim}
}

The {\verb getScatteringRecursive } program performs the actions described in \S\ref{sec:solving_a_network}, and returns the scattering matrix $\mathbf{S}$ and noise wave vector $\mathbf{c}$ for the resulting 4-port object. Nodes $\textrm{In}_{1}$, $\textrm{In}_{2}$, $\textrm{On}_{1}$, and $\textrm{On}_{2}$ are connected sequentially to ports 1 to 4 of this object.

\section{Polarimetry}

The example presented in the coming section, \S\ref{sec:arch_comparison}, compares two receiver architectures commonly used to measure linear polarization at radio to sub-mm wavelengths. This section provides necessary background information by sketching a brief summary of polarization. It shows how a receiver may be described by a Mueller matrix, and shows how to derive a receiver Mueller matrix from the scattering matrix produced by the software described in \S\ref{sec:software_implementation}.

\subsection{Brief Summary of Polarization}

An electromagnetic signal is said to be polarized if there is some lasting amplitude or phase relation between its orthogonal modes. The coherency vector \cite{Hamaker:1996p442} captures this relation:
\begin{align*}
\mathbf{e} = & \langle
\begin{bmatrix}
E_{x}(t)E_{x}^{*}(t) \\
E_{x}(t)E_{y}^{*}(t) \\
E_{y}(t)E_{x}^{*}(t) \\
E_{y}(t)E_{y}^{*}(t)
\end{bmatrix}
\rangle \\
 = & \langle \mathbf{E} \otimes \mathbf{E}^{*}  \rangle
\end{align*}
Here $\mathbf{E}$ is the complex vector of the orthogonal modes $E_{x}(t)$ and $E_{y}(t)$ of the signal, $\langle \ldots \rangle$ indicates time averaging, and $\otimes$ indicates the outer product.

If the signal $\mathbf{E}$ is acted on by an object described by a Jones matrix $\mathbf{J}$, i.e. $\mathbf{E}_{out} = \mathbf{JE}$, then the new coherency vector is given by
\begin{align}
 \mathbf{e}_{out} = & (\mathbf{J} \otimes \mathbf{J}^{*})\mathbf{e} \label{eqn:rd:coherency_vector_from_Jones}
\end{align}

The polarization state of a signal is usually described by the Stokes parameters, $I$, $Q$, $U$, and $V$. $I$ describes the total intensity of the signal, $Q$ and $U$ describe the linear polarization state, and $V$ describes the circular polarization state. The Stokes vector $\mathbf{e}^{S}$ is obtained from the coherency vector $\mathbf{e}$ by
\begin{align}
 \mathbf{e}^{S} = & 
\begin{bmatrix}
I \\ Q \\ U \\ V
\end{bmatrix}
= \mathbf{Te} \label{eqn:rd:Stokes_vector_from_coherency} \\
\textrm{where } \mathbf{T} = &
\begin{bmatrix}
 1 & 0 & 0 & 1 \\
 1 & 0 & 0 & -1 \\
 0 & 1 & 1 & 0 \\
 0 & -i & i & 0
\end{bmatrix} \label{eqn:rd:T_matrix_definition}
\end{align}
We see that $\mathbf{T}$ is a coordinate transformation of the coherency vector to the abstract Stokes frame.

The Stokes parameters are a convenient and powerful way of the describing the state of polarization of an electromagnetic signal. From Equation~\ref{eqn:rd:Stokes_vector_from_coherency} we have:
\begin{align}
\nonumber I =& \lan \vert E_{x}(t) \vert^{2} \ran + \lan \vert E_{y}(t) \vert^{2} \ran \\
\nonumber Q =& \lan \vert E_{x}(t) \vert^{2} \ran - \lan \vert E_{y}(t) \vert^{2} \ran \\
\nonumber U = & 2\lan \Re\{ E_{x}(t)E_{y}^{*}(t) \} \ran \\
\nonumber = & \lan E_{x}(t)E_{y}^{*}(t)\ran + \lan E_{x}^{*}(t)E_{y}(t) \ran \\
\nonumber V = & 2\lan \Im\{ E_{x}(t)E_{y}^{*}(t) \} \ran \\
 = & -i[ \lan E_{x}(t)E_{y}^{*}(t)\ran - \lan E_{x}^{*}(t)E_{y}(t) \ran ] \label{eqn:Stokes_parameter_definitions}
\end{align}

\paragraph{Mueller Calculus}

Suppose that a signal defined by the complex electric field vector $\mathbf{E}$ and coherency vector $\mathbf{e}$ is modified by an object described by the Jones matrix $\mathbf{J}$. From Equations~\ref{eqn:rd:coherency_vector_from_Jones} and \ref{eqn:rd:Stokes_vector_from_coherency} we see that the output signal $\mathbf{E}_{out} = \mathbf{JE}$ will be described by the Stokes vector
\begin{align*}
 \mathbf{e}^{S}_{out} = & \mathbf{T}(\mathbf{J} \otimes \mathbf{J}^{*})\mathbf{T}^{-1} \mathbf{e}^{S} \\
 = & \mathbf{M}\mathbf{e}^{S}
\end{align*}
The matrix $\mathbf{M} = \mathbf{T}(\mathbf{J} \otimes \mathbf{J}^{*})\mathbf{T}^{-1}$ is called the Mueller matrix. It represents the action of the object characterized by Jones matrix $\mathbf{J}$ in the Stokes vector space.

Mueller calculus is a matrix method for manipulating Stokes vectors. We denote the Mueller matrix elements as
\begin{align*}
 \mathbf{M} = &
\begin{bmatrix}
 M_{II} & M_{IQ} & M_{IU} & M_{IV} \\
 M_{QI} & M_{QQ} & M_{QU} & M_{QV} \\
 M_{UI} & M_{UQ} & M_{UU} & M_{UV} \\
 M_{VI} & M_{VQ} & M_{VU} & M_{VV}
\end{bmatrix}
\end{align*}

Mueller matrices are a convenient means of describing the action of an astronomical polarimeter. Of particular interest are the $M_{QI}$ and $M_{UI}$ parameters, which describe the leakage of the total intensity $I$ into the measured linear polarization vector components. Much of the radio and mm/sub-mm spectrum is only slightly linearly polarized, hence non-zero values of $M_{QI}$ and $M_{UI}$ can imply serious contamination of the measured linear polarization vector by the total intensity signal.

\subsection{Deriving Receiver Mueller Matrix}

\begin{figure}
 \centering
 \includegraphics[width=2.5in]{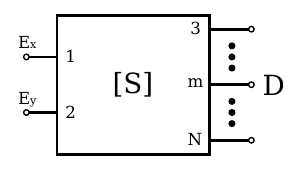}
\caption{A arbitrary receiver, where orthogonal linear polarizations $E_{x}(t)$ and $E_{y}(t)$ are presented at inputs 1 and 2 respectively. $D$ is the output at port $m$. The receiver is described by scattering matrix $\mathbf{S}$.}
 \label{fig:rd:mueller_matrix_from_scattering_matrix}
\end{figure}

Say we have calculated the scattering matrix which describes the behavior of a receiver. For polarimeters, a natural way of expressing the receiver's performance is with a Mueller matrix. We need to translate the receiver scattering matrix into a Mueller matrix which describes the action of the receiver on the Stokes vector of the incident electromagnetic signal.

Consider the arbitrary receiver shown in Figure~\ref{fig:rd:mueller_matrix_from_scattering_matrix}. Orthogonal linear polarizations $E_{x}(t)$ and $E_{y}(t)$, representing either signals in transmission lines, orthogonal modes in waveguide, or orthogonal modes in free-space, are connected to ports 1 and 2 of the receiver respectively. Receiver output $D$ is connected to port $m$. The receiver is described by the scattering matrix $\mathbf{S}$.

The output $E_{m}(t)$ seen at port $m$ is given by (assuming that the connections to ports 3 to $N$ are reflectionless):
\begin{align}
\nonumber E_m(t) = & S_{m1}E_{x}(t) + S_{m2}E_{y}(t)
\end{align}
The power contained in the signal $E_{m}(t)$ is then measured. At radio wavelengths this is often achieved through the use of a square-law detector diode. At mm and sub-mm wavelengths a bolometer might be used. The measured power $P_{D}$ is given by:
\begin{align}
\nonumber P_{D} = & \alpha \lan E_{m}(t)E_{m}(t)^{*}\ran \\
\nonumber  = & \alpha\Big[\lan \vert E_{x}(t) \vert^{2}\ran \vert S_{m1}\vert^{2} + \lan \vert E_{y}(t) \vert^{2}\ran \vert S_{m2}\vert^{2} \\
\nonumber &+ \lan E_{x}(t)E_{y}^{*}(t) \ran S_{m1}S_{m2}^{*} \\
& + \lan E_{x}^{*}(t)E_{y}(t) \ran S_{m1}^{*}S_{m2} \Big] \label{eqn:rd:P_output_scattering}
\end{align}
where $\lan \ldots \ran$ indicates time averaging, $\alpha$ is a proportionality constant dependent on the power detection method, and we assume that the instrument scattering matrix parameters are constant during the average time period. Now let
\begin{align}
\nonumber  P_{D} = & M_{DI}I + M_{DQ}Q + M_{DU}U + M_{DV}V \\
\nonumber = & M_{DI}\lan\vert E_{x}(t) \vert^{2} + \vert E_{y}(t) \vert^{2}\ran \\
\nonumber & + M_{DQ}\lan\vert E_{x}(t) \vert^{2}-\vert E_{y}(t) \vert^{2}\ran \\
\nonumber & + M_{DU}\lan E_{x}(t)E_{y}^{*}(t) +  E_{x}^{*}(t)E_{y}(t) \ran \\
& - iM_{DV}\lan E_{x}(t)E_{y}^{*}(t) - E_{x}^{*}(t)E_{y}(t) \ran \label{eqn:rd:P_output_Mueller}
\end{align}
where we have used the definition of the Stokes parameters given in Equation~\ref{eqn:Stokes_parameter_definitions}.

By comparing Equations~\ref{eqn:rd:P_output_scattering} and \ref{eqn:rd:P_output_Mueller} we can obtain the contribution of each Stokes parameter to the power measured at output D:
\begin{align}
\nonumber M_{DI} = & \frac{\alpha}{2}\{ \vert S_{m1}\vert^{2} + \vert S_{m2}\vert^{2}  \} \\
\nonumber M_{DQ} = & \frac{\alpha}{2}\{ \vert S_{m1}\vert^{2} - \vert S_{m2}\vert^{2}  \} \\
\nonumber M_{DU} = & \frac{\alpha}{2}\{ S_{m1}S_{m2}^{*} + S_{m1}^{*}S_{m2}  \} \\
 M_{DV} = & \frac{i\alpha}{2}\{ S_{m1}S_{m2}^{*} - S_{m1}^{*}S_{m2}   \}  \label{eqn:Mueller_from_S}
\end{align}
We can derive the receiver Mueller matrix by applying this technique to all the outputs of the receiver.

\section{Polarimeter Architecture Comparison} \label{sec:arch_comparison}

Two basic types of architectures are used to measure the polarization of an electromagnetic signal: differencing polarimeters, an architecture commonly used in polarization sensitive bolometer-based polarimeters \cite{Jones:2007p1886}; and pseudo-correlation (or correlation) polarimeters, an architecture commonly used in coherent, LNA or mixer based, polarimeters such as QUIET \cite{Newburgh:2010p1887}. 

Differencing polarimeters measure the difference in power between orthogonal linear modes of the electromagnetic signal; see the definition of $Q$ in Equation~\ref{eqn:Stokes_parameter_definitions} for inspiration. Correlation, or pseudo-correlation, architectures measure the polarization state by measuring the correlation between orthogonal modes. Correlation polarimeters are required to preserve the phase of the incident signal. They are hence only feasible if coherent (i.e. phase-preserving) amplifiers or mixers are available.

The choice of which architecture to use for a particular experiment is often dominated by sensitivity considerations. At low frequencies ($< \sim 60$~GHz) the availability of low-noise coherent amplifiers has favored correlation architectures \cite{Zmuidzinas:1999p1898}. At higher frequencies, the fundamental quantum limits that amplifier noise is subject to has favored direct detection technologies such as bolometers, and hence differencing polarimeter architectures, for continuum polarimetry experiments. However, continuing improvement in coherent amplifier technology at high frequencies has pushed their performance closer to the quantum limit, e.g. \cite{Gaier:2003p142}. As coherent amplifier technology improves, sensitivity may no longer be the deciding factor between technologies, and hence architectures.

Differencing and correlation architectures measure the polarization information of the incident signal in very different ways, and hence suffer from different sources of systematic error. A careful analysis of the fundamental strengths and weaknesses of each architecture is needed. {\verb SNS } is well suited to perform this analysis. In this section we use {\verb SNS } to derive receiver Mueller matrices for examples of these two polarimeter architectures. This analysis highlights the different sources of systematic error in these architectures.

\begin{figure}
 \centering
 \includegraphics[width=3.49in]{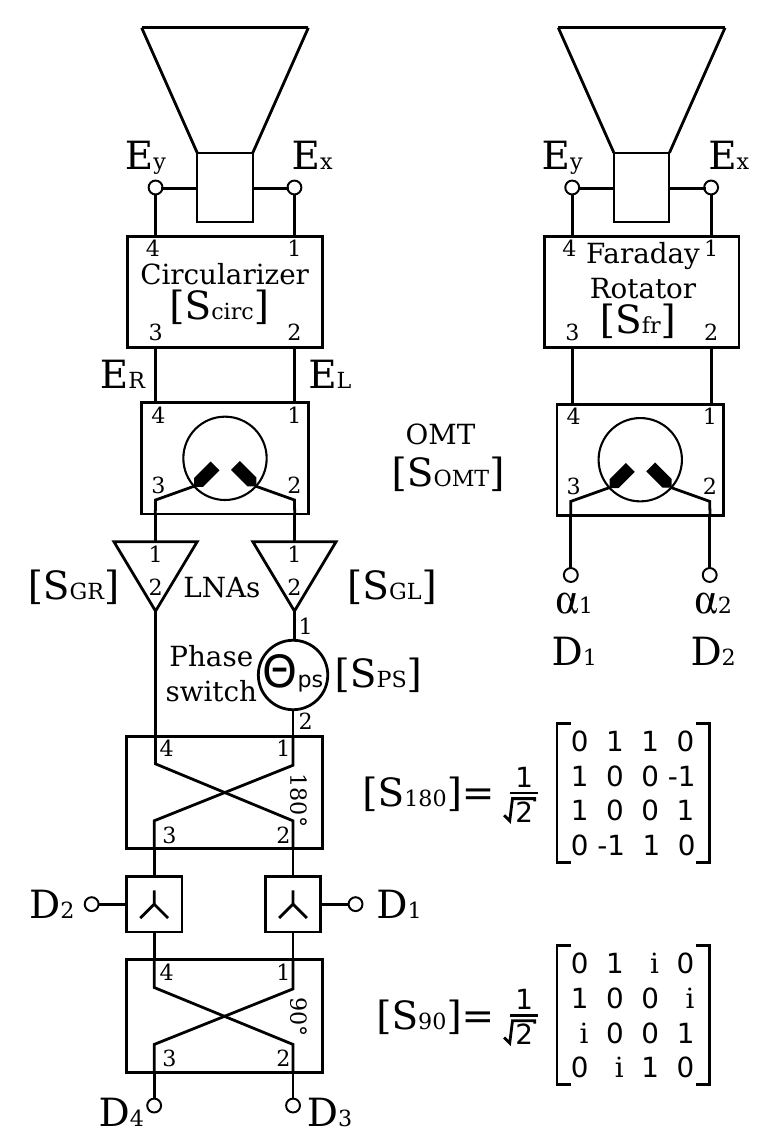}
\caption{Examples of two commonly used polarimeter architectures. (left) A pseudo-correlation polarimeter. (right) A differencing polarimeter. Orthogonal waveguide modes $E_{x}$ and $E_{y}$ are circularized or rotated, and extracted from waveguide by an orthomode transducer (OMT). In the pseudo-correlation architecture the signals are further processed. The powers in signals $D_{1}$ to $D_{4}$ are detected and processed to obtain the Stokes parameters, as explained in the text.}
 \label{fig:receiver_diagrams}
\end{figure}

\subsection{Differencing Polarimeters}
An example of a differencing polarimeter architecture is shown in Figure~\ref{fig:receiver_diagrams}~(right). The powers in orthogonal linear modes $D_{1}$ and $D_{2}$ are detected and differenced to obtain one of the linear polarization parameters. Differencing polarimeters have a much simpler architecture than pseudo-correlation polarimeters. However, they measure only a single linear Stokes parameter; a duplicate receiver oriented at $45^{\circ}$ to the first is needed to measure the second linear Stokes parameter.

\subsection{Pseudo-correlation Polarimeters}
An example of a pseudo-correlation polarimeter architecture is shown in Figure~\ref{fig:receiver_diagrams}~(left). The Stokes parameters of linear polarization in a circular basis are given by:
\begin{align}
\nonumber E_{l}(t) =& \frac{1}{\sqrt{2}}[E_{x}(t) - iE_{y}(t)] \\
\nonumber E_{r}(t) =& \frac{1}{\sqrt{2}}[E_{x}(t) + iE_{y}(t)] \\
\nonumber Q =& 2\lan \Re\{ E_{l}(t)E_{r}^{*}(t) \} \ran \\
U = & -2\lan \Im\{ E_{l}(t)E_{r}^{*}(t) \} \ran \label{eqn:rd:stokes_U_circular}
\end{align}
Pseudo-correlation polarimeters measure the linear Stokes vector by correlating circular polarization signals $E_{l}$ and $E_{r}$ with ($D_{3}$ and $D_{4}$) and without ($D_{1}$ and $D_{2}$) a $90^{\circ}$ phase shift. The powers in signals $D_{1}$ to $D_{4}$ are detected and combined to obtain the linear Stokes parameters.

While the architecture of a pseudo-correlation polarimeter is more complex than that of a differencing polarimeter, it does provide some significant advantages. For instance, both linear polarization parameters can be measured with a single optical assembly, providing twice the information for the same focal plane area occupied.

\subsection{Parameterized Scattering Matrices} \label{sec:matrix_parametrizations}
Some of the components in the receivers shown in Figure~\ref{fig:receiver_diagrams} have parameterized scattering matrices. While it is possible to describe every component in a receiver with a suitable parameterized scattering matrix, the resulting analytic expressions for the outputs soon become too complicated to be useful when written down.

In this analysis the components are assumed to be perfectly matched, i.e. the diagonal elements of the parameterized scattering matrices are zero.

\subsubsection{Circularizer}

The circular phase shifter translates orthogonal linear polarizations into orthogonal circular polarizations. It introduces a $90^{\circ}$ phase shift into one orthogonal linear mode, and is oriented at $45^{\circ}$ to the OMT linear axis.

A possible parameterization of the circularizer's scattering matrix as shown in Figure~\ref{fig:receiver_diagrams} is:
\begin{align*}
\mathbf{S}_{circ} = \frac{L_{c}}{\sqrt{2}}
\begin{bmatrix}
 0 & 1 & 1 & 0 \\
 1 & 0 & 0 & -e^{i(\frac{\pi}{2}+\theta_{c})} \\
 1 & 0 & 0 & e^{i(\frac{\pi}{2}+\theta_{c})} \\
 0 & -e^{i(\frac{\pi}{2}+\theta_{c})}  & e^{i(\frac{\pi}{2}+\theta_{c})}  & 0
\end{bmatrix} 
\end{align*}
Here $L_{c}^{2}$ is the insertion loss of the circularizer, and $\theta_{c}$ is the error in the $90^{\circ}$ phase shift. The circularizer is assumed to be otherwise perfect.

\subsubsection{Faraday Rotator}

The Faraday rotator shown in the differencing polarimeter in Figure~\ref{fig:receiver_diagrams} is a component sometimes used in PSBs \cite{Keating:2008p1850}. It modulates the measured polarization signal by introducing a variable rotation to the plane of linear polarization of the incident signal. A similar effect can be achieved with a rotating birefringent half-waveplate, or a wire grid.

The scattering matrix for the rotator considered here is:
\begin{align*}
\mathbf{S}_{fr} = 
\begin{bmatrix}
 0 & \cos(\theta_{ps}) & \sin(\theta_{ps}) & 0 \\
 \cos(\theta_{ps}) & 0 & 0 & -\sin(\theta_{ps}) \\
 \sin(\theta_{ps}) & 0 & 0 & \cos(\theta_{ps}) \\
 0 & -\sin(\theta_{ps})  & \cos(\theta_{ps})  & 0
\end{bmatrix} 
\end{align*}
Here $2\theta_{ps}$ is the time-dependent linear plane rotation introduced by the Faraday rotator.

\subsubsection{OMT}
The OMT extracts orthogonal linear modes from the waveguide. The scattering matrix parameterization considered here is:
\begin{align*}
\mathbf{S}_{OMT} = 
\begin{bmatrix}
 0 & D_{x} & d_{yx} & 0 \\
D_{x} & 0 & 0 & d_{xy} \\
d_{yx} & 0 & 0 & D_{y}\\
 0 & d_{xy} & D_{y}  & 0
\end{bmatrix} 
\end{align*}
Here $D_{x}$ and $D_{y}$ measure the insertion loss for each orthogonal mode, while $d_{xy}$ and $d_{yx}$ measure the leakage of one mode into the other. From conservation of energy considerations in a passive component we have the constraints $\vert D_{x} \vert = \vert D_{y} \vert = \abs{D}$, $\vert d_{xy} \vert = \vert d_{yx} \vert = \abs{d}$, and $\abs{D}^{2}+\abs{d}^{2}=L_{omt}^{2}$, where $L_{omt}^{2}$ is the insertion loss of the OMT. The parameters may have arbitrary phase.

\subsubsection{LNAs and Phase Switch}
The scattering matrices for the LNAs in the pseudo-correlation polarimeter are given by:
\begin{align*}
\mathbf{S}_{L,R} = 
\begin{bmatrix}
 0 &  0 \\
G_{L,R}  & 0
\end{bmatrix} 
\end{align*}
Here $G_{L}$ and $G_{R}$ are the complex voltage gains of the left and right circular polarization amplifiers respectively.

The phase switch in the pseudo-correlation polarimeter modulates the phase of one of the signal arms relative to the other. Its scattering matrix is given by:
\begin{align*}
\mathbf{S}_{ps} = 
\begin{bmatrix}
 0 &  e^{i\theta_{ps}} \\
e^{i\theta_{ps}}  & 0
\end{bmatrix} 
\end{align*}
Here $\theta_{ps}$ is the time-dependent phase shift introduced by the phase switch, usually shifting between $0^{\circ}$ and $180^{\circ}$.

\subsection{Polarimeter Mueller Matrix Elements}

We now build a connected model of each polarimeter in {\verb SNS } using the matrix parameterizations given in \S\ref{sec:matrix_parametrizations}. We use Equation~\ref{eqn:Mueller_from_S} to obtain expressions for the powers measured at each receiver output in terms of the incident signal Stokes parameters.

For the ideal pseudo-correlation polarimeter, i.e. where all the components are perfect, the outputs are:
\begin{align}
\nonumber P_{D_{1}} = & \frac{1}{2} I - \frac{1}{2} U \\
\nonumber P_{D_{2}} = & \frac{1}{2} I + \frac{1}{2} U \\
\nonumber P_{D_{3}} = & \frac{1}{2} I -\frac{1}{2} Q \\
P_{D_{4}} = & \frac{1}{2} I + \frac{1}{2} Q \label{eqn:correlation_polarimeter_diode_powers}
\end{align}
Here we have assumed that $\theta_{ps} = 0$ and $\alpha = 1$. To measure the Stokes parameters we take $Q_{m} = P_{D_{4}} - P_{D_{3}}$, $U_{m} = P_{D_{2}}- P_{D_{1}}$, and $I_{m} = \frac{1}{2}(P_{D_{1}}+P_{D_{2}}+P_{D_{3}}+P_{D_{4}})$.

For the ideal differencing polarimeter the outputs are:
\begin{align}
\nonumber P_{D_{1}} = & \frac{1}{2} I + \frac{1}{2}\cos(2\theta_{ps}) Q - \frac{1}{2}\sin(2\theta_{ps}) U \\
P_{D_{2}} = & \frac{1}{2} I - \frac{1}{2}\cos(2\theta_{ps}) Q + \frac{1}{2}\sin(2\theta_{ps}) U \label{eqn:differencing_polarimeter_diode_powers}
\end{align}
Which linear Stokes parameter we measure depends on the plane rotation introduced by the Faraday rotator, and is given by $L_{m} = P_{D_{1}}-P_{D_{2}}=\cos(2\theta_{ps}) Q  - \sin(2\theta_{ps}) U $. The measured total intensity is $I_{m} = P_{D_{1}}+P_{D_{2}}$.

The Mueller matrix parameters of particular interest in CMB polarization studies are: $M_{II}$, $M_{QQ}$, and $M_{UU}$, the diagonal elements of the Mueller matrix;  $M_{QI}$ and $M_{UI}$, the leakage of the total intensity signal into the (generally) small linear polarization signal; and $M_{QU}$ and $M_{UQ}$, which measure the rotation of the linear polarization vector by the receiver.

These Mueller matrix parameters for the pseudo-correlation polarimeter are:
{\small
\begin{align}
\nonumber M_{II} = & \frac{L_{c}^{2}L_{omt}^{2}}{2}\Big[ \abs{G_{L}}^{2} + \abs{G_{R}}^{2} \Big] \\
\nonumber M_{QQ} = & L_{c}^{2}\abs{G_{L}G_{R}}\Big[ \abs{D}^{2}\cos(\theta_{3}) + \abs{d}^{2}\cos(\theta_{4}) \Big] \\
\nonumber M_{UU} = & L_{c}^{2}\abs{G_{L}G_{R}}\Big[ \cos(\theta_{c})\{ \abs{D}^{2}\cos(\theta_{3}) - \abs{d}^{2}\cos(\theta_{4}) \} \\
\nonumber & - \sin(\theta_{c})\abs{Dd}\{\sin(\theta_{1}) + \sin(\theta_{2}) \} \Big] \\
\nonumber M_{QI} = & L_{c}^{2}\abs{G_{L}G_{R}}\abs{Dd}\Big[ \cos(\theta_{2}) +\cos(\theta_{1}) \Big] \\
\nonumber M_{UI} = & L_{c}^{2}\abs{G_{L}G_{R}}\abs{Dd}\Big[\sin(\theta_{1}) - \sin(\theta_{2}) \Big] \\
\nonumber M_{QU} = & L_{c}^{2}\abs{G_{L}G_{R}}\Big[ \cos(\theta_{c})\{ \abs{D}^{2}\sin(\theta_{3}) - \abs{d}^{2}\sin(\theta_{4}) \} \\
\nonumber & + \sin(\theta_{c})\abs{Dd}\{\cos(\theta_{2}) - \cos(\theta_{1}) \} \Big] \\
M_{UQ} = & -L_{c}^{2}\abs{G_{L}G_{R}}\Big[ \abs{D}^{2}\sin(\theta_{3}) + \abs{d}^{2}\sin(\theta_{4}) \Big] \label{eqn:correlation_pol_mueller_matrix_elements} \\
\nonumber \textrm{where } \theta_{1} = & \theta_{D_{y}} - \theta_{d_{xy}} - (\theta_{G_{L}} - \theta_{G_{R}}+\theta_{ps}) \\
\nonumber \theta_{2} = & \theta_{D_{x}} - \theta_{d_{yx}} + (\theta_{G_{L}} - \theta_{G_{R}}+\theta_{ps}) \\
\nonumber \theta_{3} = & \theta_{D_{x}} - \theta_{D_{y}} + (\theta_{G_{L}} - \theta_{G_{R}}+\theta_{ps}) \\
\nonumber \theta_{4} = & \theta_{d_{xy}} - \theta_{d_{yx}} + (\theta_{G_{L}} - \theta_{G_{R}}+\theta_{ps})
\end{align}
}
Here $X = \abs{X}e^{i\theta_{X}}$. We have implicitly assumed that the responses of all the power detectors are equal and stable.

To keep the comparison between the architectures reasonable, we need to include varying power detection sensitivity in the differencing polarimeter. Let the power detection proportionality constants (see Equation~\ref{eqn:rd:P_output_scattering}) be $\alpha_{1}$ and $\alpha_{2}$ for outputs 1 and 2 respectively. We also need to decide on the rotation angle of the Faraday rotator to specify which linear Stokes parameter we actually measure. Let $\theta_{ps} = \pm 45^{\circ}$ (i.e. Stokes U). We then obtain the Mueller matrix parameters:
{\small
\begin{align}
\nonumber M_{II} = & \frac{L_{omt}^{2}}{2}\Big[ \alpha_{1} +  \alpha_{2}\Big] \\
\nonumber M_{UU} = & \frac{\sin(2\theta_{ps})}{2}\Big[ \alpha_{1} +  \alpha_{2}\Big] \Big[ \abs{D}^{2} - \abs{d}^{2}  \Big] -\cos(2\theta_{ps})\abs{Dd} \\
\nonumber & \Big[ \alpha_{1}\cos(\theta_{D_{x}}-\theta_{d_{xy}}) -  \alpha_{2} \cos(\theta_{D_{y}}-\theta_{d_{yx}})  \Big] \\
\nonumber M_{UI} = & \frac{L_{omt}^{2}}{2}\Big[ \alpha_{1} -  \alpha_{2}\Big] \\
\nonumber M_{UQ} = & \frac{\cos(2\theta_{ps})}{2}\Big[ \alpha_{2} - \alpha_{1}\Big]\Big[ \abs{D}^{2} - \abs{d}^{2} \Big] -\sin(2\theta_{ps})\abs{Dd}\\
 & \Big[ \alpha_{1}\cos(\theta_{D_{x}}-\theta_{d_{xy}}) -  \alpha_{2} \cos(\theta_{D_{y}}-\theta_{d_{yx}})  \Big] \label{eqn:differencing_pol_mueller_matrix_elements}
\end{align}
}

\subsubsection{Discussion} \label{sec:architecture_comparison}


One of the greatest sources of systematic error in polarimeters is leakage of the total intensity signal into the measured linear polarization amplitude, $P = \sqrt{Q^{2}+U^{2}}$. The fractional contribution to $P$ from total intensity leakage, $\Delta P_{I} = M_{PI}/M_{II}$, is given by:
\begin{align*}
\Delta P_{I} = \frac{\sqrt{M_{QI}^{2}+M_{UI}^{2}}}{M_{II}} 
\end{align*}

Assume that we have two differencing polarimeters oriented such that they measure $Q$ and $U$ respectively, identical except for their values of $\alpha$. The ``$Q$'' polarimeter has values $\alpha_{1},\alpha_{2}$, while the ``$U$'' polarimeter has $\alpha_{3},\alpha_{4}$. For the pseudo-correlation and differencing polarimeters we then have:
\begin{align*}
\Delta P_{I}^{c} = & \frac{2\sqrt{2}\abs{G_{L}G_{R}}}{\abs{G_{L}}^{2}+\abs{G_{R}}^{2}}\abs{d}\sqrt{(1-\abs{d}^{2})(1+\cos(\theta_{1}+\theta_{2}))} \\
\Delta P_{I}^{d} = & 2\frac{\sqrt{(\alpha_{1}-\alpha_{2})^{2}+(\alpha_{3}-\alpha_{4})^{2}}}{\alpha_{1}+\alpha_{2}+\alpha_{3}+\alpha_{4}}
\end{align*}
As a simplification, assume that $\Delta \alpha = (\alpha_{1}-\alpha_{2})/\alpha_{2}=(\alpha_{3}-\alpha_{4})/\alpha_{4}$, and let $\Delta \abs{G}^{2} = \frac{\abs{G_{L}}^{2}-\abs{G_{R}}^{2}}{\abs{G_{R}}^{2}}$. Take the worst-case phase scenario, where $\cos(\theta_{1}+\theta_{2})=1$. We now have:
\begin{align}
\Delta P_{I}^{c} = & 4\abs{d}\sqrt{1-\abs{d}^{2}}\frac{\sqrt{1+\Delta \abs{G}^{2}}}{2+\Delta \abs{G}^{2}} \label{eqn:corr_ItoP_leakage} \\
\Delta P_{I}^{d} = & \frac{\sqrt{2}\Delta \alpha}{2+\Delta \alpha} \label{eqn:diff_ItoP_leakage}
\end{align}

\begin{figure}
 \centering
 \includegraphics[width=3.25in]{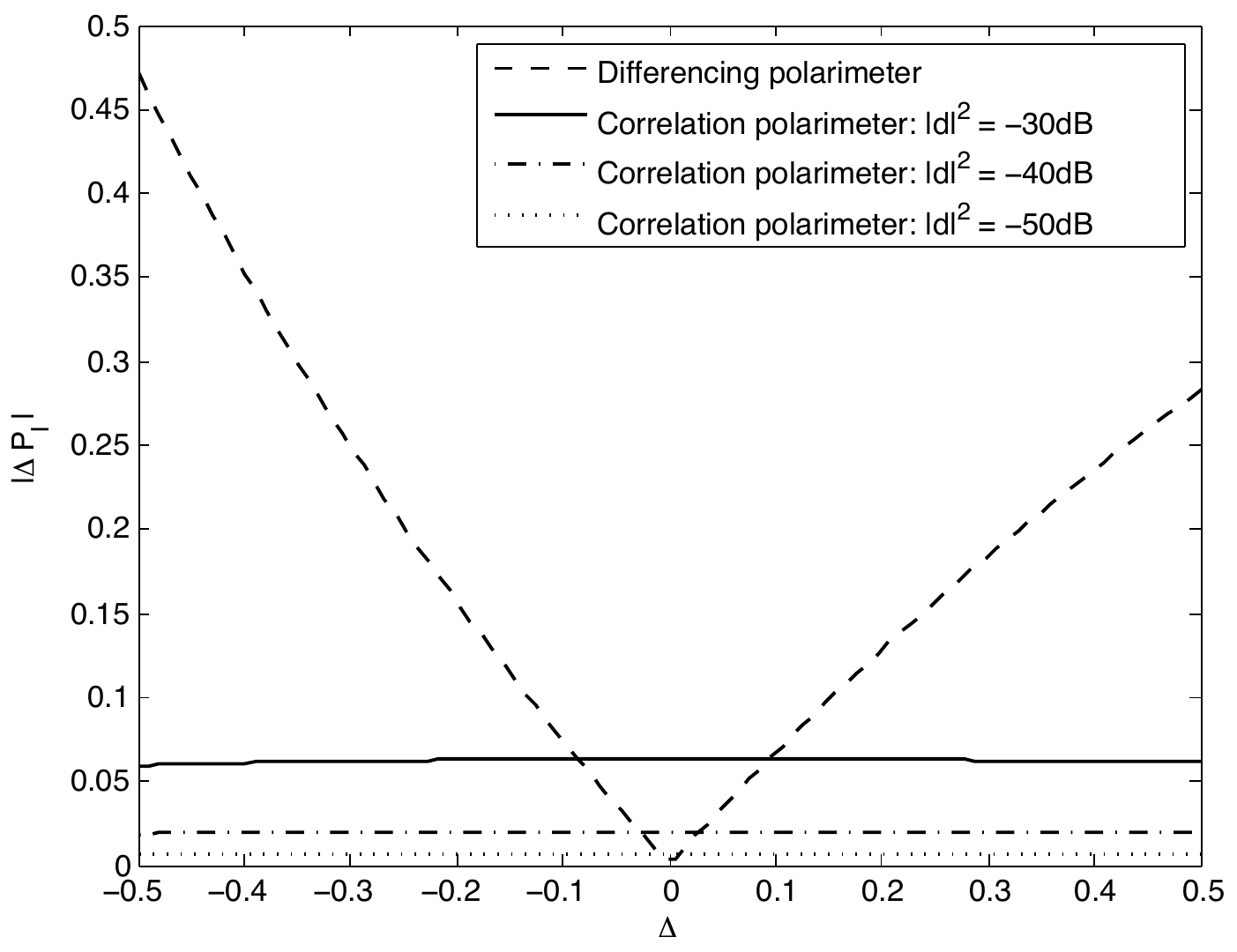}
\caption{Comparison of the effect of imbalance on the fractional total intensity to polarization leakage for pseudo-correlation and differencing polarimeter architectures. For the pseudo-correlation architecture, $\Delta = \Delta\abs{G}^{2}$. For the differencing architecture $\Delta = \Delta\alpha$. $\abs{d}^{2}$ is the OMT cross polarization.}
 \label{fig:ItoPleakage_comparison}
\end{figure}

Equations~\ref{eqn:corr_ItoP_leakage} and \ref{eqn:diff_ItoP_leakage} are plotted in Figure~\ref{fig:ItoPleakage_comparison}. Several attributes are noteworthy: $\Delta P_{I}$ is \emph{independent} of the OMT cross polarization $\abs{d}^{2}$ for the differencing polarimeter, but is heavily dependent on the power sensitivity imbalance $\Delta \alpha$; $\Delta P_{I}$ is almost independent of gain imbalance $\Delta \abs{G}^{2}$ for the pseudo-correlation polarimeter, but is dependent on the OMT cross polarization.

Figure~\ref{fig:ItoPleakage_comparison} clearly illustrates the difference between the polarimeter architectures in terms of total intensity to polarization leakage. Correlation polarimeters are very  insensitive to what is generally the most unstable parameter in a coherent receiver: fluctuating LNA gain. They are moderately sensitive to OMT cross polarization $\abs{d}^{2}$. The comparatively high sensitivity of differencing polarimeters to power detection imbalance can be reduced if $\Delta\alpha$ is stable and well-known; the data can then be corrected and the leakage of $I$ into $P$ reduced.

\subsection{Pseudo-Correlation Polarimeter Noise Temperature}

A very powerful benefit of using scattering matrices to model receivers is the ability to perform noise analysis. We specify parameterized noise wave vectors for the pre-LNA components and ignore any noise produced by the components ``down stream'' of the LNAs, as their contribution will be negligible if the LNA gain is high.

If the noise wave vector of the pseudo-correlation polarimeter is given by $\mathbf{c}$, then the noise power measured at the output $D_{i}$ in a 1~Hz bandwidth is given by $P_{i} = \alpha \lan c_{j}c_{j}^{*}\ran$, where $c_{j}$ is the noise wave vector element corresponding to output $D_{i}$. 

Suppose that component $k$ of $M$ total components has $N$ ports, and is specified by the scattering matrix $\mathbf{S}^{k}$ and the noise wave vector $\mathbf{c}^{k}$. $c_{j}$ is given by:
\begin{align*}
c_{j} = & \sum_{k=1}^{M} c_{j}^{k}, \textrm{where } c_{j}^{k} = \sum_{i=1}^{N}b_{i}^{k}c_{i}^{k}
\end{align*}

Noise waves from different devices are not correlated: $\lan c_{i}^{k}(c_{j}^{m})^{*} \ran = 0$ for $k\neq m$. So, $P_{i}$ is given by:
\begin{align*}
P_{i} = & \alpha \sum_{k=1}^{M}P_{i}^{k} \\
\textrm{where }P_{i}^{k} = & \sum\sum\Big( \mathbf{C}^{k}\cdot \big(\mathbf{b}^{k} \otimes (\mathbf{b}^{k})^{\dagger}\big) \Big)
\end{align*}
Here $\mathbf{C}^{k}$ is the noise correlation matrix for component $k$, $\cdot$ is the matrix dot product, $\mathbf{b}^{k}$ is the vector $[b_{1}^{k} \ldots b_{N}^{k}]^{\textrm{T}}$, $\otimes$ is the outer product, and $^{\dagger}$ is the hermitian conjugate. The $\sum\sum$ indicates a sum over all the matrix elements.

The noise correlation matrices for the passive pre-LNA components are obtained using Equation~\ref{eqn:noise_correlation_matrix_for_passive_device}. To get the noise correlation matrices for the LNAs we make two simplifications to the HEMT noise correlation matrix model in \cite{Wedge:1992p149}: first, the off-diagonal terms of an LNA noise correlation matrix ($\lan c_{1}c_{2}^{*}\ran$ and $\lan c_{1}^{*}c_{2}\ran$) are much smaller than the diagonal terms so we take them to be zero. Second: $\lan \vert c_{2} \vert^{2} \ran \simeq \vert S_{21} \vert^{2} \lan \vert c_{1} \vert^{2} \ran \simeq k \vert S_{21} \vert^{2}T_{\textrm{N}}$, where $T_{\textrm{N}}$ is the amplifier noise temperature and $k$ is Boltzmann's constant. 

Receiver noise temperature is referenced to the input. We consider the receiver temperature $T_{i}$ at output $D_{i}$ to be the temperature of a thermal source seen equally at each input which produces the same total output noise power in a noiseless receiver:
\begin{align}
\nonumber P_{i} = & \alpha \Big( \sum_{j=1}^{N_{\textrm{in}}}\abs{S_{ij}}^{2} \Big) T_{i} \\
\therefore T_{i} = & \frac{\sum_{k=1}^{M} P_{i}^{k}}{\sum_{j=1}^{N_{\textrm{in}}}\abs{S_{ij}}^{2}}
\end{align}
Here $N_{\textrm{in}}$ is the number of inputs to the receiver, and $\mathbf{S}$ is the receiver scattering matrix.

Applying this technique to the pseudo-correlation polarimeter we derive the receiver temperatures for the outputs $D_{1}$ to $D_{4}$:
\begin{align}
\nonumber T_{1} = & T_{c} + \frac{T_{p}}{L_{c}^{2}}\Big[ \frac{G}{E-F} - 1 \Big] + \frac{T_{N}}{L_{c}^{2}}\Big[ \frac{G}{E-F}\Big] \\
\nonumber T_{2} = & T_{c} + \frac{T_{p}}{L_{c}^{2}}\Big[ \frac{G}{E+F} - 1 \Big] + \frac{T_{N}}{L_{c}^{2}}\Big[ \frac{G}{E+F}\Big] \\
\nonumber T_{3} = & T_{c} + \frac{T_{p}}{L_{c}^{2}}\Big[ \frac{G}{E-H} - 1 \Big] + \frac{T_{N}}{L_{c}^{2}}\Big[ \frac{G}{E-H}\Big] \\
T_{4} = & T_{c} + \frac{T_{p}}{L_{c}^{2}}\Big[ \frac{G}{E+H} - 1 \Big] + \frac{T_{N}}{L_{c}^{2}}\Big[ \frac{G}{E+H}\Big] \label{eqn:T_expressions} \\
\textrm{where }\nonumber T_{c} = & T_{p}\Big[\frac{1}{L_{c}^{2}}-1\Big] \\
\nonumber G = & \abs{G_{L}}^{2}+\abs{G_{R}^{2}} \\
\nonumber E = & \big(\abs{G_{L}}^{2}+\abs{G_{R}^{2}}\big)L_{omt}^{2} \\
\nonumber F = & 2\abs{G_{L}G_{R}}\abs{Dd}\Big[ \cos(\theta_{1}) + \cos(\theta_{2}) \Big] \\
\nonumber H = & 2\abs{G_{L}G_{R}}\abs{Dd}\Big[\sin(\theta_{1}) - \sin(\theta_{2}) \Big]
\end{align}
Here we have assumed that the amplifiers have noise temperature $T_{N}$, and that the circularizer and OMT are at a physical temperature of $T_{p}$. $\theta_{1}$ and $\theta_{2}$ are given in Equation~\ref{eqn:correlation_pol_mueller_matrix_elements}.

\subsubsection{Discussion}

Which parameters in the multiparameter expression for $T_{1}$ in Equation~\ref{eqn:T_expressions} have the greatest impact on the receiver temperature? The most obvious are $L_{c}^{2}$ and $L_{omt}^{2}$, the insertion losses of the circularizer and OMT respectively. Setting those aside, how do the other parameters affect the receiver temperature?

The minimum value that $T_{1}$ can have is $T_{N}$. Our sensitivity impact metric is then $T_{1}/T_{N} \geq 1$. We fix $L_{c}^{2} = -0.1$~dB and $L_{omt}^{2}=-0.2$~dB, and set $\abs{G_{R}}=1$, $\abs{G_{L}}=\sqrt{1+\Delta\abs{G}^{2}}$, $\theta_{G_{L}}=0$, and $\theta_{D_{x}}=0$ (only relative phases matter here). This leaves us with six parameters: $\abs{d}^{2}$, $\Delta\abs{G}^{2}$, $\theta_{G_{R}}$, $\theta_{D_{y}}$, $\theta_{d_{xy}}$, and $\theta_{d_{yx}}$.

\begin{figure}
 \centering
 \includegraphics[width=3.49in]{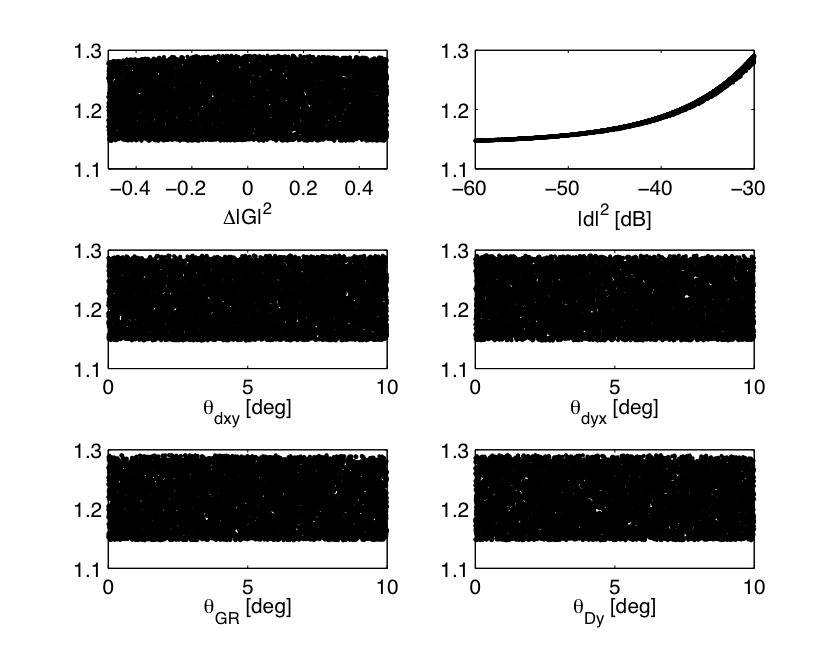}
\caption{Plots of $T_{1}/T_{N}$ against various parameters. We see that the receiver is most sensitive to the cross polarization, $\abs{d}^{2}$, and is negligibly sensitive to the other parameters. We assume $T_{N}=T_{P}=15$~K. See text for details.}
 \label{fig:T3_sensitivity}
\end{figure}

We generate random sets of physically realistic values for these parameters, and evaluate the metric $T_{1}/T_{N}$ for each set. We plot $T_{1}/T_{N}$ against each parameter under consideration in Figure~\ref{fig:T3_sensitivity}. It is immediately clear that the most important parameter in this set is $\abs{d}^{2}$. Discarding the least important parameters we now have:
\begin{align}
\nonumber T_{1} \backsimeq &  \frac{1}{L_{c}^{2}}\frac{T_{p}+T_{N}}{L_{omt}^{2}-2\abs{d}\sqrt{L_{omt}^{2}-\abs{d}^{2}}} - T_{p} \\
 \backsimeq & \frac{T_{p}(1-L_{c}^{2}L_{omt}^{2})+T_{N}}{L_{c}^{2}L_{omt}^{2}} \textrm{ since } \abs{d}^{2} \ll L_{omt}^{2} \label{eqn:T1_simplified}
\end{align}
This simplified expression for $T_{1}$ is exactly what we would derive using a conventional noise temperature analysis, indicating that the software has calculated the noise temperature correctly.

\section{Conclusions}

We have presented {\verb SNS }, a MATLAB-based software library written to aid in the design and analysis of receiver architectures. It uses electrical scattering matrices and noise wave vectors to describe receiver architectures of arbitrary topology and complexity.

We use {\verb SNS } to compare two polarimeter architectures commonly used to perform measurements of the polarized CMB: differencing polarimeters, an architecture commonly used in PSB-based polarimeters; and pseudo-correlation polarimeters, an architecture commonly used in coherent polarimeters. This analysis highlights the differing sources of systematic error in these architectures: $I$ to $P$ leakage in pseudo-correlation polarimeters is almost immune to gain imbalance, but sensitive to OMT cross polarization; while $I$ to $P$ leakage in differencing polarimeters is immune to OMT cross polarization, but very sensitive to power detection imbalance.

We show how {\verb SNS } can be used to calculate analytical expressions for the receiver noise temperature of arbitrary receivers. Analytic expressions for the receiver temperature of a pseudo-correlation polarimeter are derived, and are found to be consistent with those obtained from conventional receiver temperature calculations.

\section*{Acknowledgments}

Thanks go to Dr Paul Grimes for helpful discussions and pointers to papers, to Prof. Mike Jones for helpful suggestions, support and motivation, and to Nikolai Yu. Zolotykh for developing the {\verb pointer } library for MATLAB. OGK acknowledges the support of a Dorothy Hodgkin Award in funding his studies while a student at Oxford, and the support of a W.M. Keck Institute for Space Studies Postdoctoral Fellowship at Caltech.

\end{document}